\documentclass[conference]{IEEEtran}
\IEEEoverridecommandlockouts
\usepackage{cite}
\usepackage{amsmath,amssymb,amsfonts}
\usepackage{algorithmic}
\usepackage{graphicx}
\usepackage{textcomp}
\usepackage{xcolor}
\def\BibTeX{{\rm B\kern-.05em{\sc i\kern-.025em b}\kern-.08em
    T\kern-.1667em\lower.7ex\hbox{E}\kern-.125emX}}
\begin{document}

\title{Discrete Optimal Transport and Voice Conversion 
\thanks{The code is available at https://anton-selitskiy.github.io/dotvc/ Published in ICLMT 2026.}
}

\author{\IEEEauthorblockN{Anton Selitskiy}
\IEEEauthorblockA{\textit{Electric and Computer Engineering Dept.} \\
\textit{University of Rochester}\\
Rochester (NY), USA \\
aselitsk@ur.rochester.edu}
\and
\IEEEauthorblockN{ Maitreya Kocharekar}
\IEEEauthorblockA{\textit{Computer Science Dept.} \\
\textit{Rochester Institute of Technology}\\
Rochester (NY), USA \\
mk1651@g.rit.edu}
}

\maketitle

\begin{abstract}
We propose $k$DOT, a discrete optimal transport (OT) framework for voice conversion (VC) operating in a pretrained speech embedding space. In contrast to the averaging strategies used in $k$NN-VC and SinkVC, and the independence assumption adopted in MKL, our method employs the barycentric projection of the discrete OT plan to construct a transport map between source and target speaker embedding distributions.

We conduct a comprehensive ablation study over the number of transported embeddings and systematically analyze the impact of source and target utterance duration. Experiments on LibriSpeech demonstrate that OT with barycentric projection consistently improves distribution alignment and often outperforms averaging-based approaches in terms of WER, MOS, and FAD.

Furthermore, we show that applying discrete OT as a post-processing step can transform spoofed speech into samples that are misclassified as bona fide by a state-of-the-art spoofing detector. This demonstrates the strong domain adaptation capability of OT in embedding space, while also revealing important security implications for spoof detection systems.
\end{abstract}


\section{Introduction}
\label{sec:intro}

Voice conversion (VC) aims to modify a speech signal from a source speaker so that it is perceived as spoken by a target speaker, while preserving the underlying linguistic content. Recent advances in self-supervised speech models have enabled VC systems that operate directly in high-dimensional embedding spaces extracted from pretrained encoders such as wav2vec, HuBERT, and WavLM. 

Notably, one of the training objectives of WavLM~\cite{wavlm} includes speaker identification, which encourages the model to retain speaker-specific information within its embeddings. As a result, these representations capture not only linguistic content but also speaker identity, making the explicit extraction of attributes such as fundamental frequency (F0) unnecessary.

Previous VC methods rely on local averaging strategies. For example, $k$NN-VC \cite{VCNN} maps each source embedding to the average of its $k$ nearest neighbors in the target set. A refinement \cite{VCOT} using discrete optimal transport (DOT) replaces nearest neighbors with assignments obtained from a Sinkhorn transport plan. However, prior work fixes the number of transported embeddings to $k=4$ and relies on heuristic averaging of selected targets.

In this work, we revisit discrete and continuous OT for VC. Instead of averaging selected target embeddings, we use the barycentric projection of the OT plan to define a transport map between empirical speaker distributions. This yields a mathematically grounded transformation that remains well-defined even when all target embeddings are used.

Our contributions are:

\begin{itemize}
    \item We introduce an OT-based barycentric projection ($k$DOT) method for voice conversion in pretrained embedding space.
    \item We perform a systematic ablation study over the number of transported embeddings and analyze the role of source and target duration.
    \item We demonstrate that $k$DOT improves distribution alignment as measured by FAD and achieves competitive or superior WER and MOS.
    \item We reveal that discrete OT can act as a powerful adversarial domain-alignment mechanism. 
\end{itemize}

Our results show that discrete OT provides an effective and theoretically grounded framework for embedding-level voice conversion, while also exposing important security considerations.

\section{Optimal Transport}
\label{sec:2}

\subsection{Discrete Optimal Transport}

Let $(X,\mathcal{P},\mathbb{P})$ and $(Y,\mathcal{Q},\mathbb{Q})$ be two probability spaces. Denote by $\Pi(\mathbb{P},\mathbb{Q})$ all joint distributions on the product space $(X\times Y, \mathcal{P}\otimes\mathcal{Q}, \pi)$ with marginals $\mathbb{P}$ and $\mathbb{Q},$ i.e., 
$\pi(A\times Y) = \mathbb{P}(A)$ for all $A\in\mathcal{P}$ and $\pi(X\times B) = \mathbb{Q}(B)$ for all $B\in\mathcal{Q}.$ The goal of optimal transport (OT) is to find the joint distribution $\pi\in \Pi(\mathbb{P},\mathbb{Q})$ known as \textit{Kantorovich plan,}   that minimizes the expected transport cost
\begin{equation}\label{otk}
\int\limits_{X\times Y} c(x,y) d\pi(x,y) \to \underset{\pi\in \Pi(\mathbb{P},\mathbb{Q})}{\inf},
\end{equation}
where $c(x,y)$ is a cost function.

In \textit{discrete case} assume  there are $M$ vectors in $X$ and $N$ vectors in $Y$ with probability masses $p_i=\mathbb{P}(x_i)$ and $q_j=\mathbb{Q}(y_j).$ Then the joint distribution $\pi(x,y)$ is represented as a non-negative matrix $\gamma$ with $\gamma_{ij}=\pi(x_i,y_j),$ $i=1,\ldots,M$ and $j=1,\ldots,N.$ The objective~(\ref{otk}) becomes:
\begin{equation}
    \sum_{i}^{M} \sum_{j}^{N} \gamma_{ij} c(x_i,y_j) \to \underset{\gamma_{ij}}{\inf},
\end{equation}
subject to the marginal constraints:
\begin{equation}
    p_i = \sum_{j=1}^{N} \gamma_{ij}\quad \text{and}\quad q_j = \sum_{i=1}^{M} \gamma_{ij}.
\end{equation}
Given a solution $\gamma$, a transport map can be defined via the \textit{barycentric projection:}
\begin{equation}\label{bar}
    T(x_i) = \sum_{j=1}^{N} \tilde{\gamma}_{ij}y_j, \quad \text{where } \tilde{\gamma}_{ij} = \frac{\gamma_{ij}}{p_i}.
\end{equation}
This can be interpret as $\mathbb{E}[y|x=x_i]$ the conditional expectation.  To compute the plan $\gamma,$  we use entropic OT with Sinkhorn algorithm described in~\cite[Ch.~4]{ot:book}.

\subsection{Continuous Optimal Transport}
\label{sec:2.2}

In case of continuous distributions $\mathbb{P}$ and $\mathbb{Q},$ the Monge formulation seeks  a measurable transformation $T\colon X\to Y$ with $\mathbb{P}\circ T^{-1} = \mathbb{Q}$ that minimizes the transport cost
\begin{equation}\label{monge}
    \int\limits_{X} c(x,T(x)) d\mathbb{P}(x) \to \underset{T}{\inf}.
\end{equation}

Under mild conditions, the solution of the Kantorovich optimal transport (OT) problem can be derived from the solution of the Monge problem; 
however, this relationship does not hold in the discrete setting. In particular, the map $T$ in~(\ref{bar}) is not the exact solution to problem~(\ref{monge}) --- which requires a deterministic map --- but rather an approximation. When paired examples are available, for instance from DOT, the \textit{flow matching} method can be employed to compute an approximate continuous transport map $T$~\cite[Th.~4.2]{flowm}. For neural network-based approximations of the transport map, see~\cite{korotin2023neural}.


\section{Voice Conversion Algorithm}
\label{sec:3}

\subsection{Interface} We use WavLM Large pretrained model~\cite{wavlm} to encode  every $25$ ms of audio into a $1024$-dimensional vector embedding, with a hop size of $20$ ms. 
For every pair of audio recordings $(x,y)$ from the source and target speakers respectively, we extract their vectorized representations:
\begin{equation}
\mathbf{x}=[x_1,\ x_2,\ldots, x_M],\quad \mathbf{y}=[y_1,\ y_2,\ldots, y_N],
\end{equation}
where $x_i, y_j\in \mathbb{R}^{1024}.$ We transform $\mathbf{x}\mapsto \mathbf{\hat{y}}=[\hat{y}_1,\ldots, \hat{y}_M].$

\subsection{Marginal distributions}

Since the underlying distributions of speaker embeddings $X=\{x_i\}_{i=1}^{M}$ and $Y=\{y_j\}_{j=1}^{N}$ are unknown, we  use  \textit{empirical distributions:} 
\begin{equation}\label{marg}
\mathbb{P}(x_i)=\frac{1}{M}\quad \text{and}\quad \mathbb{P}(y_j)=\frac{1}{N}. 
\end{equation}

\subsection{Cost function}

While the standard  cost function in OT is the $\ell_2$ distance, for high-dimensional vector embeddings,   cosine similarity is often more appropriate. Because we want the smaller cost for more similar vectors, we define the cost function as: 
\begin{equation}\label{cost}
    c(x,y) = 1 - \cos(x,y).
\end{equation}

\subsection{Transportation Map}

\paragraph{$k$NN-VC Mapping} In $k$NN-VC approach~\cite{VCNN}, $\hat{y}_i$ is determined as regression of $x_i$ on $Y,$ with the cosine similarity distance.

\paragraph{SinkVC Mapping} In the discrete OT approach, we compute the  matrix $\gamma$ for the  marginal distributions~(\ref{marg}) and cost function~(\ref{cost}). For each $x_i,$ we sort the target embeddings $y_j$ in decreasing order of $\gamma_{ij},$  denoting  the sorted vectors as $y^{sort(i)}_j.$ The sorted coupling weights along each row (with fixed $i$) are denoted  by $\gamma^{sort}_{ij}.$

In paper~\cite{VCOT}, each source embedding $x_i$ is mapped to an average of four target embeddings corresponding to top-4 $\gamma_{i1},\ldots,\gamma_{iN}$
This method, referred to as \emph{SinkVC} in~\cite{VCOT}, derives its name from the Sinkhorn algorithm commonly used to solve DOT problems.

\paragraph{MKL Mapping} To reduce computational cost, \cite{got} proposed to use the closed-form Gaussian OT map. In our experiments, this approach performs poorly, which may explain the introduction of a variance-stratified scheme in~\cite{got}. In that scheme, the standard deviations of individual embedding components
$
\sigma_k = \operatorname{std}(x_{1k}, \ldots, x_{Mk}),$ $k=1,\ldots,1024,
$
are computed, and the embedding dimensions are sorted in decreasing order of $\sigma_k$. The reordered embeddings $\tilde{x}_i$ and $\tilde{y}_j$ are then partitioned into blocks of equal size $B$, and continuous OT is applied independently to each block. The authors consider $B \in \{2, 8, 16, 256\}$ and refer to this method as \emph{factorized MKL} (Monge---Kantorovich Linear). While this approach significantly improves computational efficiency and can be applied to short utterances, treating embedding components as uncorrelated across blocks leads to a noticeable loss of target speaker identity (see Table~\ref{tab:fad}).

\paragraph{$k$DOT Mapping} We define the $k$DOT mapping as the truncated baricentric projection of the OT map over top-$k$ vectors,
\begin{equation}
    x_i \overset{T}{\mapsto} \hat{y}_i  = \sum_{j=1}^{k} \tilde{\gamma}^{sort}_{ij}y^{sort(i)}_j,\quad \tilde{\gamma}^{sort}_{ij} = \frac{\gamma^{sort}_{ij}}{\sum_{s=1}^{k}\gamma^{sort}_{is}}.
\end{equation}
In contrast to $k$NN-VC and SinkVC, this formulation remains well-defined even for $k = N$. When $k = 1,$ SinkVC and $k$DOT coincide. In practice, however, using all target embeddings may lead to noisy outputs due to the uniform marginal assumption, as many embeddings correspond to silence or low-energy segments. To improve robustness, we therefore restrict the summation to the top-$k$ transport weights.


\subsection{Vocoder}

After the transformation $\mathbf{x}\mapsto \mathbf{\hat{y}},$ we convert  the predicted embeddings $\mathbf{\hat{y}}$ back into waveform $\hat{y}$ using HiFi-GAN vocoder.

\section{Experiments and Evaluation}
\label{sec:4}

\subsection{Voice Conversion on LibriSpeech}
\label{sec:4.1}

We conduct our experiments using the LibriSpeech train-clean-100  dataset~\cite{panayotov2015librispeech}. Following the protocol in~\cite{VCOT}, we select the first 40 speakers (ordered by speaker ID) and, for each, extract 10 random utterances and sort them by duration. Each speaker is converted into the voice of the remaining 39 speakers.

To investigate the impact of audio duration on VC performance, we evaluate the following cases: 

\begin{enumerate}
\item Cumulative duration less than 5 seconds --- typically includes one or no utterances per speaker.
\item Cumulative duration less than 1 minute --- typically includes 2–3 utterances per speaker.
\item All 10 utterances --- typically results in a cumulative duration of approximately 100 seconds.
\end{enumerate}

For each experimental case, we conduct an ablation study over different values of $k$. The Word Error Rate (WER)~\cite{wer} and Mean Opinion Score (MOS)~\cite{mos} results are reported in Tables~\ref{tab1}--\ref{tab3}. We provide standard deviations; since 100 samples were used, the corresponding 95\% confidence intervals can be obtained by multiplying the reported deviations by 0.2. 

Additionally, for Case~3 (the full set of utterances), we report the Fr\'echet Audio Distance (FAD)~\cite{fad} in Table~\ref{tab:fad}, as this metric provides a more reliable estimate when a sufficiently large number of embeddings is available.


\begin{table}[htbp]
\caption{Case 1. Source and target are shorter than 5 sec}
\begin{center}
\setlength{\tabcolsep}{3pt}
\begin{tabular}{lccccc}
\hline
  & $k$NN & SinkVC & $k$DOT & MKL-2 & MKL-256\\
  \hline
  \multicolumn{6}{c}{WER$\downarrow$}\\
  \hline
$k=1$ & $1.05\pm 0.19$ & $\mathbf{0.96\pm 0.09}$ & $\mathbf{0.96\pm 0.09}$ &\rotatebox{90}{$\mathbf{0}$}& \rotatebox{90}{$5$}\\
$k=3$ & $0.78\pm 0.09$ & $0.79\pm 0.07$ & $\mathbf{0.77\pm 0.08}$ &\rotatebox{90}{$\mathbf{2}$}&\rotatebox{90}{$8$}\\
$k=4$ & $1.27\pm 0.79$ & $1.27\pm 0.79$ & $\mathbf{0.76\pm 0.09}$ &\rotatebox{90}{$\mathbf{0.}$}& \rotatebox{90}{$1.$}\\
$k=5$ & $\mathbf{0.69\pm 0.08}$ & $0.75\pm 0.16$ & $0.74\pm 0.07$ &\rotatebox{90}{$\mathbf{\pm}$}& \rotatebox{90}{$\pm$}\\
$k=10$ & $\mathbf{0.67\pm 0.08}$ & $0.70\pm 0.08$ & $0.69\pm 0.07$ &\rotatebox{90}{$\mathbf{9}$}& \rotatebox{90}{$1$}\\
$k=40$ & $\mathbf{0.65\pm 0.07}$ & $0.76\pm 0.08$ & $0.71\pm 0.09$ &\rotatebox{90}{$\mathbf{2}$}& \rotatebox{90}{$9$}\\ 
$k=N$ & --- & --- & $\mathbf{0.91\pm 0.07}$ & \rotatebox{90}{$\mathbf{0.}$} & \rotatebox{90}{$0.$}\\
\hline
\multicolumn{6}{c}{MOS$\uparrow$}\\
\hline
$k=1$ & $2.31\pm 0.11$ & $\mathbf{2.37\pm 0.11}$ & $\mathbf{2.37\pm 0.11}$ &\rotatebox{90}{$\mathbf{2}$}& \rotatebox{90}{$9$}\\
$k=3$ & $2.28\pm 0.11$ & $\mathbf{2.51\pm 0.13}$ & $\mathbf{2.51\pm 0.10}$ &\rotatebox{90}{$\mathbf{5}$}&\rotatebox{90}{$6$}\\
$k=4$ & $\mathbf{2.49\pm 0.12}$ & $2.20\pm 0.11$ & ${2.45\pm 0.12}$ &\rotatebox{90}{$\mathbf{0.}$}& \rotatebox{90}{$1.$}\\
$k=5$ & $\mathbf{2.44\pm 0.11}$ & $2.42\pm 0.11$ & $2.38\pm 0.12$ &\rotatebox{90}{$\mathbf{\pm}$}& \rotatebox{90}{$\pm$}\\
$k=10$ & $\mathbf{2.30\pm 0.10}$ & $2.15\pm 0.11$ & $2.18\pm 0.13$ &\rotatebox{90}{$\mathbf{6}$}& \rotatebox{90}{$6$}\\
$k=40$ & $\mathbf{1.93\pm 0.11}$ & $1.87\pm 0.10$ & $1.80\pm 0.13$ &\rotatebox{90}{$\mathbf{8}$}& \rotatebox{90}{$2$}\\ 
$k=N$ & --- & --- & $\mathbf{1.93\pm 0.11}$ & \rotatebox{90}{$\mathbf{2.}$} & \rotatebox{90}{$2.$}\\

\hline
\end{tabular}
\label{tab1}
\end{center}
\end{table}

\begin{table}[htbp]
\caption{Case 2. Source and target are shorter than 1 min}
\begin{center}
\setlength{\tabcolsep}{3pt}
\begin{tabular}{lccccc}
\hline
  & $k$NN & SinkVC & $k$DOT & MKL-2 & MKL-256\\
  \hline
  \multicolumn{6}{c}{WER$\downarrow$}\\
  \hline
$k=1$ & ${0.29\pm 0.03}$ & $\mathbf{0.28\pm 0.03}$ & $\mathbf{0.28\pm 0.03}$ &\rotatebox{90}{$\mathbf{2}$}& \rotatebox{90}{$\mathbf{1}$}\\
$k=3$ & $\mathbf{0.25\pm 0.04}$ & $\mathbf{0.25\pm 0.02}$ & $0.26\pm 0.03$ &\rotatebox{90}{$\mathbf{1}$}&\rotatebox{90}{$\mathbf{1}$}\\
$k=4$ & $\mathbf{0.24\pm 0.03}$ & $0.25\pm 0.04$ & $0.26\pm 0.04$ &\rotatebox{90}{$\mathbf{0.}$}& \rotatebox{90}{$\mathbf{0.}$}\\
$k=5$ & ${0.25\pm 0.03}$ & ${0.25\pm 0.03}$ & $\mathbf{0.23\pm 0.02}$ &\rotatebox{90}{$\mathbf{\pm}$}& \rotatebox{90}{$\mathbf{\pm}$}\\
$k=10$ & $\mathbf{0.23\pm 0.02}$ & $0.25\pm 0.03$ & $\mathbf{0.23\pm 0.03}$ &\rotatebox{90}{$\mathbf{0}$}& \rotatebox{90}{$\mathbf{0}$}\\
$k=40$ & $\mathbf{0.23\pm 0.03}$ & $0.24\pm 0.04$ & $0.25\pm 0.03$ &\rotatebox{90}{$\mathbf{2}$}& \rotatebox{90}{$\mathbf{2}$}\\ 
$k=N$ & --- & --- & $\mathbf{0.35\pm 0.04}$ & \rotatebox{90}{$\mathbf{0.}$} & \rotatebox{90}{$\mathbf{0.}$}\\
\hline
\multicolumn{6}{c}{MOS$\uparrow$}\\
\hline
$k=1$ & $2.59\pm 0.10$ & $\mathbf{2.75\pm 0.12}$ & $\mathbf{2.75\pm 0.12}$ &\rotatebox{90}{$8$}& \rotatebox{90}{$5$}\\
$k=3$ & $2.90\pm 0.15$ & $\mathbf{2.92\pm 0.12}$ & $\mathbf{2.92\pm 0.10}$ &\rotatebox{90}{${5}$}&\rotatebox{90}{$5$}\\
$k=4$ & ${2.97\pm 0.13}$ & $\mathbf{2.98\pm 0.12}$ & ${2.96\pm 0.11}$ &\rotatebox{90}{${0.}$}& \rotatebox{90}{$1.$}\\
$k=5$ & $\mathbf{3.00\pm 0.13}$ & $2.87\pm 0.11$ & $2.90\pm 0.11$ &\rotatebox{90}{${\pm}$}& \rotatebox{90}{$\pm$}\\
$k=10$ & $\mathbf{2.98\pm 0.12}$ & $2.93\pm 0.11$ & $2.87\pm 0.13$ &\rotatebox{90}{$0$}& \rotatebox{90}{$0$}\\
$k=40$ & ${2.63\pm 0.13}$ & $\mathbf{2.69\pm 0.14}$ & $2.67\pm 0.11$ &\rotatebox{90}{${8}$}& \rotatebox{90}{$6$}\\ 
$k=N$ & --- & --- & $\mathbf{1.78\pm 0.11}$ & \rotatebox{90}{${2.}$} & \rotatebox{90}{$2.$}\\

\hline
\end{tabular}
\label{tab2}
\end{center}
\end{table}


\begin{table}[htbp]
\caption{Case 3. Source and target are longer than 1 min}
\begin{center}
\setlength{\tabcolsep}{3pt}
\begin{tabular}{lccccc}
\hline
  & $k$NN & SinkVC & $k$DOT & MKL-2 & MKL-256\\
  \hline
  \multicolumn{6}{c}{WER$\downarrow$}\\
  \hline
$k=1$ & $0.26\pm 0.03$ & $\mathbf{0.25\pm 0.03}$ & $\mathbf{0.25\pm 0.03}$ &\rotatebox{90}{$2$}& \rotatebox{90}{$3$}\\
$k=3$ & $\mathbf{0.21\pm 0.02}$ & $0.24\pm 0.03$ & ${0.22\pm 0.02}$ &\rotatebox{90}{${1}$}&\rotatebox{90}{$1$}\\
$k=4$ & $\mathbf{0.22\pm 0.03}$ & $0.23\pm 0.02$ & ${0.23\pm 0.02}$ &\rotatebox{90}{${0.}$}& \rotatebox{90}{$0.$}\\
$k=5$ & ${0.25\pm 0.03}$ & $0.24\pm 0.03$ & $\mathbf{0.23\pm 0.02}$ &\rotatebox{90}{${\pm}$}& \rotatebox{90}{$\pm$}\\
$k=10$ & $\mathbf{0.22\pm 0.02}$ & $\mathbf{0.22\pm 0.02}$ & $\mathbf{0.22\pm 0.02}$ &\rotatebox{90}{${2}$}& \rotatebox{90}{$2$}\\
$k=40$ & ${0.24\pm 0.03}$ & $\mathbf{0.21\pm 0.02}$ & $0.24\pm 0.02$ &\rotatebox{90}{${2}$}& \rotatebox{90}{$2$}\\ 
$k=N$ & --- & --- & $\mathbf{0.36\pm 0.05}$ & \rotatebox{90}{${0.}$} & \rotatebox{90}{$0.$}\\
\hline
\multicolumn{6}{c}{MOS$\uparrow$}\\
\hline
$k=1$ & $\mathbf{2.85\pm 0.13}$ & ${2.84\pm 0.12}$ & ${2.84\pm 0.12}$ &\rotatebox{90}{${3}$}& \rotatebox{90}{$6$}\\
$k=3$ & $2.97\pm 0.11$ & ${2.99\pm 0.10}$ & $\mathbf{3.04\pm 0.12}$ &\rotatebox{90}{${5}$}&\rotatebox{90}{$5$}\\
$k=4$ & $3.01\pm 0.09$ & $\mathbf{3.14\pm 0.13}$ & ${2.98\pm 0.13}$ &\rotatebox{90}{${0.}$}& \rotatebox{90}{$0.$}\\
$k=5$ & ${3.02\pm 0.10}$ & ${3.12\pm 0.13}$ & $\mathbf{3.13\pm 0.12}$ &\rotatebox{90}{${\pm}$}& \rotatebox{90}{$\pm$}\\
$k=10$ & $\mathbf{3.06\pm 0.12}$ & $2.99\pm 0.11$ & $3.04\pm 0.11$ &\rotatebox{90}{${5}$}& \rotatebox{90}{$4$}\\
$k=40$ & $\mathbf{2.91\pm 0.14}$ & $2.69\pm 0.13$ & $2.88\pm 0.12$ &\rotatebox{90}{${8}$}& \rotatebox{90}{$7$}\\ 
$k=N$ & --- & --- & $\mathbf{1.88\pm 0.16}$ & \rotatebox{90}{${2.}$} & \rotatebox{90}{$2.$}\\

\hline
\end{tabular}
\label{tab3}
\end{center}
\end{table}

To further disentangle the role of source and target duration, we consider two additional asymmetric cases:

\begin{enumerate}
\item[4)] Source duration $>$ 1 minute, target duration $<$ 1 minute (Table~\ref{tab4}).
\item[5)] Source duration $<$ 1 minute, target duration $>$ 1 minute (Table~\ref{tab5}).
\end{enumerate}



\begin{table}[htbp]
\caption{Case 4. Source longer 1 min, target shorter 1 min}
\begin{center}
\setlength{\tabcolsep}{3pt}
\begin{tabular}{lccccc}
\hline
  & $k$NN & SinkVC & $k$DOT & MKL-2 & MKL-256\\
  \hline
  \multicolumn{6}{c}{WER$\downarrow$}\\
  \hline
$k=1$ & $\mathbf{0.68\pm 0.05}$ & ${0.74\pm 0.08}$ & $0.74\pm 0.08$ &\rotatebox{90}{${4}$}& \rotatebox{90}{$\mathbf{1}$}\\
$k=3$ & $0.80\pm 0.47$ & $0.57\pm 0.04$ & $\mathbf{0.52\pm 0.04}$ &\rotatebox{90}{${1}$}&\rotatebox{90}{$\mathbf{1}$}\\
$k=4$ & $0.54\pm 0.05$ & $\mathbf{0.51\pm 0.04}$ & ${0.53\pm 0.05}$ &\rotatebox{90}{${0.}$}& \rotatebox{90}{$\mathbf{0.}$}\\
$k=5$ & $0.50\pm 0.05$ & $\mathbf{0.48\pm 0.04}$ & $0.50\pm 0.04$ &\rotatebox{90}{${\pm}$}& \rotatebox{90}{$\mathbf{\pm}$}\\
$k=10$ & $0.46\pm 0.05$ & $0.46\pm 0.04$ & $\mathbf{0.41\pm 0.04}$ &\rotatebox{90}{${1}$}& \rotatebox{90}{$\mathbf{9}$}\\
$k=40$ & $0.48\pm 0.04$ & $0.54\pm 0.14$ & $\mathbf{0.45\pm 0.05}$ &\rotatebox{90}{${2}$}& \rotatebox{90}{$\mathbf{1}$}\\ 
$k=N$ & --- & --- & $\mathbf{0.64\pm 0.08}$ & \rotatebox{90}{${0.}$} & \rotatebox{90}{$\mathbf{0.}$}\\
\hline
\multicolumn{6}{c}{MOS$\uparrow$}\\
\hline
$k=1$ & ${2.39\pm 0.12}$ & $\mathbf{2.42\pm 0.09}$ & $\mathbf{2.42\pm 0.09}$ &\rotatebox{90}{$\mathbf{1}$}& \rotatebox{90}{$0$}\\
$k=3$ & $\mathbf{2.51\pm 0.11}$ & ${2.24\pm 0.12}$ & $\mathbf{2.34\pm 0.11}$ &\rotatebox{90}{$\mathbf{5}$}&\rotatebox{90}{$5$}\\
$k=4$ & $\mathbf{2.31\pm 0.10}$ & $2.25\pm 0.10$ & ${2.24\pm 0.10}$ &\rotatebox{90}{$\mathbf{0.}$}& \rotatebox{90}{$0.$}\\
$k=5$ & ${2.24\pm 0.09}$ & $\mathbf{2.48\pm 0.13}$ & $2.25\pm 0.12$ &\rotatebox{90}{$\mathbf{\pm}$}& \rotatebox{90}{$\pm$}\\
$k=10$ & ${2.17\pm 0.11}$ & $\mathbf{2.24\pm 0.12}$ & $2.22\pm 0.10$ &\rotatebox{90}{$\mathbf{7}$}& \rotatebox{90}{$3$}\\
$k=40$ & ${1.97\pm 0.12}$ & $\mathbf{2.06\pm 0.12}$ & $1.77\pm 0.13$ &\rotatebox{90}{$\mathbf{8}$}& \rotatebox{90}{$7$}\\ 
$k=N$ & --- & --- & $\mathbf{1.79\pm 0.12}$ & \rotatebox{90}{$\mathbf{2.}$} & \rotatebox{90}{$2.$}\\

\hline
\end{tabular}
\label{tab4}
\end{center}
\end{table}


\begin{table}[htbp]
\caption{Case 5. Source shorter 1 min, target longer than 1 min}
\begin{center}
\setlength{\tabcolsep}{3pt}
\begin{tabular}{lccccc}
\hline
  & $k$NN & SinkVC & $k$DOT & MKL-2 & MKL-256\\
  \hline
  \multicolumn{6}{c}{WER$\downarrow$}\\
  \hline
$k=1$ & $0.37\pm 0.05$ & $\mathbf{0.36\pm 0.04}$ & $\mathbf{0.36\pm 0.04}$ &\rotatebox{90}{$\mathbf{9}$}& \rotatebox{90}{$5$}\\
$k=3$ & $\mathbf{0.32\pm 0.05}$ & $0.38\pm 0.05$ & ${0.35\pm 0.04}$ &\rotatebox{90}{$\mathbf{1}$}&\rotatebox{90}{$5$}\\
$k=4$ & $0.36\pm 0.04$ & $0.33\pm 0.04$ & $\mathbf{0.32\pm 0.04}$ &\rotatebox{90}{$\mathbf{0.}$}& \rotatebox{90}{$1.$}\\
$k=5$ & $\mathbf{0.32\pm 0.04}$ & $0.33\pm 0.04$ & $\mathbf{0.32\pm 0.04}$ &\rotatebox{90}{$\mathbf{\pm}$}& \rotatebox{90}{$\pm$}\\
$k=10$ & $\mathbf{0.31\pm 0.04}$ & $0.34\pm 0.05$ & $\mathbf{0.31\pm 0.04}$ &\rotatebox{90}{$\mathbf{8}$}& \rotatebox{90}{$1$}\\
$k=40$ & ${0.30\pm 0.04}$ & $0.31\pm 0.04$ & $\mathbf{0.29\pm 0.04}$ &\rotatebox{90}{$\mathbf{2}$}& \rotatebox{90}{$3$}\\ 
$k=N$ & --- & --- & $\mathbf{0.63\pm 0.08}$ & \rotatebox{90}{$\mathbf{0.}$} & \rotatebox{90}{$1.$}\\
\hline
\multicolumn{6}{c}{MOS$\uparrow$}\\
\hline
$k=1$ & $\mathbf{2.85\pm 0.12}$ & $2.71\pm 0.12$ & $2.71\pm 0.12$ &\rotatebox{90}{${0}$}& \rotatebox{90}{$2$}\\
$k=3$ & $\mathbf{3.09\pm 0.12}$ & ${2.92\pm 0.13}$ & $3.02\pm 0.11$ &\rotatebox{90}{${5}$}&\rotatebox{90}{$5$}\\
$k=4$ & $2.91\pm 0.14$ & $3.07\pm 0.11$ & $\mathbf{3.09\pm 0.12}$ &\rotatebox{90}{${0.}$}& \rotatebox{90}{$0.$}\\
$k=5$ & $2.95\pm 0.11$ & $\mathbf{3.17\pm 0.13}$ & $2.98\pm 0.11$ &\rotatebox{90}{${\pm}$}& \rotatebox{90}{$\pm$}\\
$k=10$ & $2.88\pm 0.12$ & $\mathbf{3.12\pm 0.11}$ & $3.01\pm 0.12$ &\rotatebox{90}{$7$}& \rotatebox{90}{$5$}\\
$k=40$ & ${2.91\pm 0.13}$ & $2.84\pm 0.14$ & $\mathbf{2.94\pm 0.11}$ &\rotatebox{90}{$6$}& \rotatebox{90}{$6$}\\ 
$k=N$ & --- & --- & $\mathbf{1.66\pm 0.12}$ & \rotatebox{90}{${2.}$} & \rotatebox{90}{$2.$}\\

\hline
\end{tabular}
\label{tab5}
\end{center}
\end{table}

\begin{table}[ht]
\centering
\caption{FAD $\downarrow$ in case source and target are longer than 1 min}
\label{tab:fad}
\setlength{\tabcolsep}{4pt}
\begin{tabular}{lcccccc}
    \hline
    & {$k=1$} & {$k=3$} & {$k=4$} & {$k=5$} & {$k=10$} & {$k=40$} \\
    \hline
    $k$NN-VC\cite{VCNN}  &  1.148 & 0.826 & 0.807 & 0.815 & \bfseries 0.878 & 1.410\\
    SinkVC\cite{VCOT}   &    \bfseries 1.150 & 0.824 & 0.789 & \bfseries 0.789 & 0.890 & 1.415\\
    $k$DOT   &    \bfseries 1.150 & \bfseries 0.787 & \bfseries 0.780 & 0.796 & 0.884 & \bfseries 1.394\\
    MKL-2\cite{got} & \multicolumn{6}{c}{$1.173$}\\
    MKL-256\cite{got} & \multicolumn{6}{c}{$1.389$}\\
    \hline
\end{tabular}
\end{table}

\paragraph{Impact of Duration.}

Tables~\ref{tab1}--\ref{tab3} reveal a clear dependency of VC performance on the cumulative duration of available utterances. When both source and target are extremely short (Case~1), performance is unstable across all methods. Increasing duration to 1 minute (Case~2) dramatically reduces WER (from $\approx 1.0$ to $\approx 0.25$) and improves MOS by nearly one point. When both sides exceed one minute (Case~3), performance saturates and becomes more consistent across $k$ values.

The asymmetric setups (Cases~4 and~5) further clarify this behavior. Comparing Tables~\ref{tab4} and~\ref{tab5}, we observe that increasing \emph{target} duration has a substantially larger effect than increasing source duration. In particular, Case~5 (short source, long target) consistently achieves lower WER and higher MOS than Case~4 (long source, short target). This confirms that the richness of the target embedding distribution is the dominant factor for perceptual quality and intelligibility, consistent with earlier observations in~\cite{VCNN,VCOT}.

\paragraph{Effect of $k$ and Over-Smoothing.}

Across all duration regimes, moderate values of $k$ (typically $k=3$--$5$) provide the best trade-off between intelligibility and naturalness. Very small $k$ values ($k=1$) lead to high variance and unstable alignment, while very large $k$ (especially $k=N$) consistently degrade both WER and MOS. 

This behavior indicates an over-smoothing phenomenon: when too many embeddings are averaged or transported, the resulting representation drifts toward the barycenter of the target distribution, reducing speaker distinctiveness and harming intelligibility. 


\paragraph{Comparison of Methods.}

The proposed $k$DOT method achieves competitive or superior WER across most values of $k$, particularly in Cases~4 and~5, where domain imbalance is present. For all methods except MKL, performance differences across moderate values of $k$ generally fall within the reported confidence intervals, preventing definitive conclusions about the model choice. Importantly, $k$DOT consistently yields lower Fr\'echet Audio Distance (FAD) in the high-duration setting (Table~\ref{tab:fad}), indicating improved global distribution alignment.

Interestingly, the MKL variants perform competitively in short-duration settings but exhibit higher FAD, indicating that they tend to preserve certain source characteristics rather than fully align with the target distribution. Subjectively, this results in speech that remains perceptually closer to the source speaker despite improvements in local reconstruction metrics. This behavior suggests that WavLM embeddings encode speaker identity in a distributed and correlated manner: components with higher variance appear more strongly associated with clarity and intelligibility, while full distribution alignment requires structured transport rather than independent component manipulation.

Finally, since these methods do not rely on task-specific pretraining for particular languages, they can be readily applied to underrepresented or low-resource languages. 




\paragraph{Distribution Alignment vs. Intelligibility.}

The FAD results (Table~\ref{tab:fad}) show that $k$DOT achieves the best alignment to the target distribution for moderate $k$ values. However, improvements in FAD do not always directly translate to lower WER or higher MOS. This highlights an important distinction: global distribution matching and local intelligibility are related but not identical objectives. Effective voice conversion requires balancing both.

\subsection{Voice Conversion on ASVspoof 2019}
\label{sec:5}

To evaluate OT under severe domain mismatch, we conduct experiments on the ASVspoof 2019 dataset~\cite{wang2020asvspoof}. Based on earlier duration analysis, we retain only utterances longer than 2.9 seconds. We convert the first 1000 spoofed recordings into the last 1000 bona fide recordings using a one-to-one OT setup. The converted signals are then evaluated using the AASIST spoof detection model~\cite{Jung2021AASIST}. 

To isolate the effect of OT from potential vocoder artifacts, we include an encode-decode control condition (WavLM Large + HiFi-GAN) applied to both bona fide and spoofed signals without transport. This reconstruction introduces mild degradation but does not significantly affect spoof detection accuracy.

\begin{figure}[ht]
  \centering
  \includegraphics[width=\columnwidth]{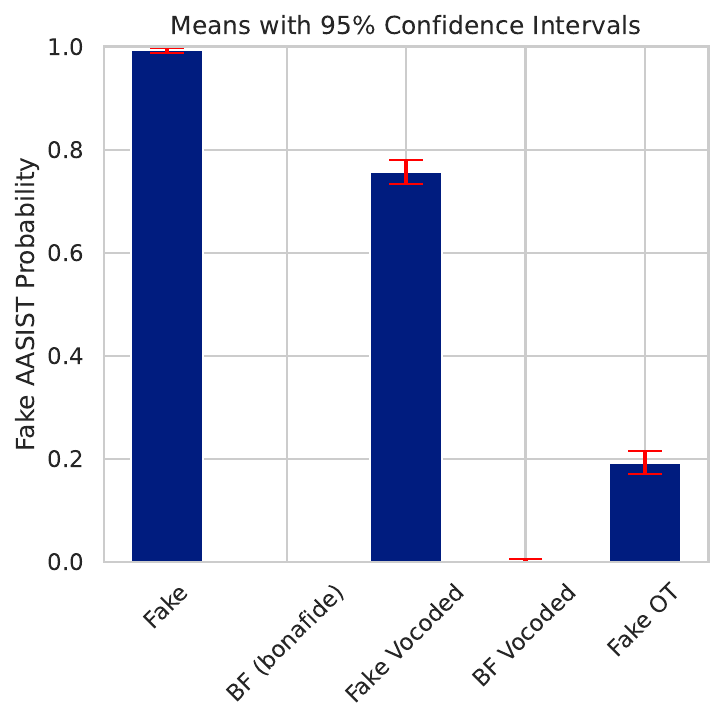}
  \vspace{-2em}
  \caption{AASIST probabilities of fake audio.}
  \label{fig:aasist}
\end{figure}

In contrast, applying discrete OT induces a dramatic distribution shift: more than 80\% of spoofed samples are classified as bona fide (Fig.~\ref{fig:aasist}). This corresponds to a substantial increase in Equal Error Rate (Table~\ref{tab:eer}). 

These results indicate that $k$DOT does not merely smooth embeddings but actively aligns spoofed and bona fide distributions in a way that removes discriminative artifacts. From a security perspective, this reveals that embedding-level optimal transport can function as a powerful adversarial domain-alignment mechanism capable of bypassing state-of-the-art spoof detection systems.





\begin{table} 
\centering
\caption{Equal Error Rate (EER$\downarrow$) for strongest spoofing attacks}
\label{tab:eer}
\begin{tabular}{ccc}
    \hline
    A18& A18$_{5}$ & 4DOT  \\
    \hline
    $2.614$  &  $0.435$  &   $11.111$ \\
    \hline
\end{tabular}
\end{table}

\section{Experimental Setup}
\label{sec:6}
\subsection{Datasets}

For the voice conversion experiments (Sec.~\ref{sec:4.1}), we use the LibriSpeech Clean 100 subset~\cite{kaggleLS}. 

For spoofing experiments (Sec.~\ref{sec:5}), we use ASVspoof 2019~\cite{wang2020asvspoof} and evaluate against the strongest A18 attack as well as Malafide (A18$_5$) from ASVspoof5 dataset~\cite{wang2024asvspoof}.



\subsection{Model Pipeline}

Speech embeddings are extracted using WavLM Large (sixth transformer layer). Waveforms are reconstructed using the HiFi-GAN vocoder from the KNN-VC repository~\cite{VCOT}.

Discrete OT is computed via the Sinkhorn algorithm using the POT library~\cite{pot} with a default entropic regularization parameter $\varepsilon=0.1$.

Spoof detection is performed using the official AASIST implementation~\cite{Jung2021AASIST}.





\subsection{Evaluation Metrics}

Word Error Rate (WER) is computed using Whisper~\cite{whisper} for transcription and JiWER for alignment.

Mean Opinion Score (MOS) is estimated automatically using UTMOSv2~\cite{utmosv2}.

Fr\'echet Audio Distance (FAD) is computed using VGGish embeddings~\cite{vgg}. Following~\cite{fadtk}, we extract pre-activation embeddings, compute mean and covariance statistics, and evaluate Fr\'echet distance between distributions.

Equal error rate (EER), i.e., the operating point where the false acceptance rate equals the false rejection rate. 




\section{Conclusion}
\label{sec:7}

We demonstrated that discrete optimal transport provides a principled and effective framework for embedding-level voice conversion. The proposed barycentric OT mapping achieves strong distribution alignment and often outperforms averaging-based approaches, particularly when sufficient target duration is available. 

We showed that moderate values of $k$ yield the best trade-off between intelligibility and naturalness, while excessive averaging leads to over-smoothing. 

Importantly, we demonstrate that OT can act as a powerful domain-alignment mechanism capable of enabling spoofed speech to bypass a state-of-the-art detector. This finding underscores both the effectiveness of $k$DOT for distribution alignment and the broader security implications of applying optimal transport in speech representation spaces.







\section{Acknowledgment}
\label{sec:ack}

The first author thanks Mona Udasi for discussions related to the results in Sec.~\ref{sec:5}, Meiying Chen for pointing to the automatic MOS evaluation method~\cite{utmosv2}, and Research Computing at the Rochester Institute of Technology for providing computational resources that supported this research.

\bibliographystyle{IEEEtran}
\bibliography{refs25}

\end{document}